\documentclass[superscriptaddress,amsmath, nofootinbib]{revtex4}
\usepackage{amsfonts}
\usepackage{graphicx}
\usepackage[latin1]{inputenc}
\usepackage{amsmath}
\usepackage{amssymb}
\begin{document}


\title{Lagrange top: integrability according to Liouville and examples of analytic solutions.}

\author{Alexei A. Deriglazov }
\email{alexei.deriglazov@ufjf.br} \affiliation{Depto. de Matem\'atica, ICE, Universidade Federal de Juiz de Fora,
MG, Brazil}

\date{\today}

\begin{abstract}
Equations of a heavy rotating body with one fixed point can be deduced starting from a variational problem with holonomic constraints. When applying this formalism to the particular case of a Lagrange top, in the formulation  with a diagonal inertia tensor the potential energy has more complicated form as compared with that assumed in the literature on dynamics of a rigid body. This implies the corresponding improvements in equations of motion. Therefore, we revised this case, presenting several examples of analytical solutions to the improved equations. The case of precession without nutation has a surprisingly rich relationship between the rotation and precession rates, and this is discussed in detail. 
\end{abstract}

\maketitle 



\section{Introduction.}   

The recent work \cite{AAD23} was devoted to a systematic exposition of the dynamics of a free rigid body, considered as  a system with holonomic (that is velocity-independent)  constraints. The constraints in the action functional have been taken into account with use of Lagrangian multipliers. Having accepted this expression, we no longer need any additional postulates or assumptions about the behavior of the rigid body. As was shown in \cite{AAD23}, all the basic quantities and characteristics of a rigid body, as well as the equations of motion and integrals of motion, are obtained from the variational problem by direct and unequivocal calculations within the framework of standard methods of classical mechanics applied in Laboratory system. Here we follow the same scheme to deduce  equations of motion of a rigid body with fixed point in the gravity field. The analysis is a similar to that presented in Sects. II - VI of the work \cite{AAD23}, so we only outline it in Sect. II and III below, wihout entering into the details.  

Then we concentrate on the case of Lagrange top. When formulating its variational problem with the diagonal inertia tensor, we observed that  potential energy has more complicated form as compared with that assumed in the literature on dynamics of a rigid body. This implies the corresponding improvements in equations of motion. Because this is a somewhat surprising observation, its validity and comparison with the literature are detailly carried out at the end of Sect. II and in Sect. IV. Being one of the classical problems of non linear dynamics and integrable systems, this issue however is of interest in the modern studies related with construction and behavior of spinning particles and rotating bodies in external fields beyond the pole-dipole approximation \cite{Abd_23, Off_23, Chak_23, AAD_2023_9, Wei_2024, Kos_2023, Ver_2024, Fri_2023, Ale_2024}. 

Using the Liouville's theorem, integration of improved equations can be reduced to the calculation of four elliptic integrals \cite{AAD23_5}. Of course, the answer in the form of elliptic integrals is not very illuminating. Therefore, in subsequent sections V-VIII we present several examples of solutions to the improved equations in terms of elementary functions: sleeping top, horizontally precessing top, as well as the inclined top precessing without nutation. For the latter case, the solution turn out to be  two-frequency motion with surprisingly reach relationship between the frequences, that, besides the inclination, depends also on the top's geometry.  We also discussed the case of an awakened top, for which there is no longer a solution in elementary functions. The qualitative and numerical analysis of this case is based on the study of effective potential.

\section{Rigid body with a fixed point.}\label{FP}

In this section we confirm that equations of motion for the rotational degrees of freedom of a rigid body with fixed point formally coincide with those of a free rigid body (see Eqs. (\ref{fp9}) below). The only difference is that all quantities (including the inertia tensor, its eigenvalues and eigenvectors) should be calculated in the Laboratory system with origin in the fixed point instead of the center of mass.

Rigid body is considered as a system\footnote{We use the notation adopted in \cite{AAD23}. In particular, 
the notation for the scalar product is:  $({\bf a}, {\bf b})=a_i b_i$. Notation for the vector product: $[{\bf a}, {\bf b}]_i=\epsilon_{ijk}a_j b_k$, 
where $\epsilon_{ijk}$ is Levi-Chivita symbol in three dimensions, with $\epsilon_{123}=+1$.} composed of $n$ points with the coordinates $y_N^i$, and masses $m_N$, $N=1, 2, \ldots , n, i=1, 2, 3$. To the constraints $({\bf y}_N(t)-{\bf y}_K(t),  {\bf y}_P(t)-{\bf y}_M(t))=\mbox{const}.$, determining a free rigid body, we add more constraints: $|{\bf y}_N(t)-{\bf y}_0|=\mbox{const}$, where ${\bf y}_0\in\mathbb{R}^3$ is some selected point.  So we have a body with the fixed point \cite{Poin,Whit_1917,Mac_1936,Lei_1965,Landau_8,Arn_1,Gol_2000,Grei_2003, Ham_22}. We place the origin of the Laboratory system at the point ${\bf y}_0$, and denote the resulting coordinates ${\bf x}_N(t)$. Then the constraints are
$({\bf x}_N(t)-{\bf x}_K(t),  {\bf x}_P(t)-{\bf x}_M(t))=\mbox{const}$, $|{\bf x}_N(t)|=\mbox{const}$, this implies  
$({\bf x}_N(t), {\bf x}_K(t))=\mbox{const}$. 
From them we can separate $3(N-1)$ independent constraints as follows. Let's take three linearly independent vectors ${\bf x}_A=({\bf x}_1, {\bf x}_2, {\bf x}_3)$ among ${\bf x}_N$, and denote others by ${\bf x}_\alpha$, $\alpha=4, 5, \ldots , n$. Let's consider all the constraints which 
involve ${\bf x}_A$
\begin{eqnarray}\label{fp2}
({\bf x}_A, {\bf x}_B)=a_{AB}=\mbox{const}, \qquad ({\bf x}_A, {\bf x}_\alpha)=a_{A\alpha}=\mbox{const}. 
\end{eqnarray}
They imply that the body has three degrees of freedom, that is the configuration space is the three-dimesional 
surface  ${\mathbb S}^3\subset\mathbb{R}^{3n}$ specified by the equations (\ref{fp2}). The Lagrangian action that takes into account these constraints is (the matrix $\lambda_{AB}$ was chosen to be the symmetric matrix):  
\begin{eqnarray}\label{fp2.1}
S=\int dt ~\frac12\sum_{N=1}^{n}m_N\dot{\bf x}_N^2+\frac12\sum_{A, B=1}^{3}\lambda_{AB}\left[({\bf x}_A, {\bf x}_B)-a_{AB}\right]+
\sum_{A=1}^{3}\sum_{\beta=4}^{n}\lambda_{A\beta}\left[({\bf x}_A, {\bf x}_\beta)-a_{A\beta}\right]. 
\end{eqnarray}
Variation of this action with respect 
to $\lambda_{AN}$ implies the constraints (\ref{fp2}), while the variation with respect to ${\bf x}_N(t)$ gives the dynamical equations
\begin{eqnarray}\label{fp3}
m_A\ddot{\bf x}_A=\sum_{B=1}^3 \lambda_{AB}{\bf x}_B+\sum_{\beta=4}^n \lambda_{A\beta}{\bf x}_\beta, \qquad 
m_\alpha\ddot{\bf x}_\alpha=\sum_{A=1}^3 \lambda_{A\alpha}{\bf x}_A. 
\end{eqnarray}
In turn, they imply the conservation of energy and angular momentum
\begin{eqnarray}\label{fp4}
\frac{dE}{dt}=0, \qquad \mbox{where} \qquad E=\frac12 \sum_{N=1}^nm_{N}\dot{\bf x}_N^2,  
\end{eqnarray}
\begin{eqnarray}\label{fp5}
\frac{d{\bf m}}{dt}=0, \qquad \mbox{where} \qquad {\bf m}=\sum_{N=1}^{n}m_N[{\bf x}_N, \dot{\bf x}_N]. 
\end{eqnarray}
Similarly to Sect. III of the work \cite{AAD23}, the constraints (\ref{fp2}) imply that any solution ${\bf x}_N(t)$, $N=1, 2, \ldots , n$ to the equations of motion is of the form 
\begin{eqnarray}\label{fp6}
{\bf x}_N(t)=R(t){\bf x}_N(0), \qquad \mbox{this implies} \qquad R_{ij}(0)=\delta_{ij}, 
\end{eqnarray}
with the same orthogonal matrix $R_{ij}(t)$ for all $N$.  Both columns and rows of the matrix $R_{ij}$ have a geometric interpretation. The 
columns $R(t)=({\bf R}_1(t), {\bf R}_2(t), {\bf R}_3(t))$ form an orthonormal basis rigidly connected to the body. The initial data for $R_{ij}(t)$, pointed in (\ref{fp6}), imply that at $t=0$ these columns coincide with the basis vectors ${\bf e}_i$ of the Laboratory system.  The rows, $R^T(t)=({\bf G}_1(t), {\bf G}_2(t), {\bf G}_3(t))$, represent the laboratory basis vectors ${\bf e}_i$ in the body-fixed basis. For example, the functions ${\bf G}_3(t)=(R_{31}(t), R_{32}(t), R_{33}(t))$ are components of ${\bf e}_3$ in the basis ${\bf R}_i(t)$. 

Assuming that ${\bf x}_N(t)$, $N=1, 2, \ldots , n$ is a solution to the equations of motion (\ref{fp3}), we can introduce the same basic characteristics that were used for the description of a free body. They are: angular 
velocity $\omega_k\equiv-\frac12 \epsilon_{kij}(\dot R R^T)_{ij}$, angular velocity in the body $\Omega_i$ defined by ${\boldsymbol\omega}(t)=\omega_i(t){\bf e}_i=\Omega_i(t){\bf R}_i(t)$, angular momentum $m_i$, and angular momentum in the body $M_i$. The relationships among them are as follows: 
\begin{eqnarray}\label{fp7}
{\bf\Omega}=R^T{\boldsymbol\omega}=I^{-1}R^T{\bf m}=I^{-1}{\bf M}. 
\end{eqnarray}
Besides, the kinetic energy (the first term in Eq. (\ref{fp2.1})) can be presented through these quantities 
\begin{eqnarray}\label{fp7.1}
E=\frac12\sum_{N=1}^{n}m_N\dot{\bf x}_N^2=
\frac12g_{ij}\dot{\bf R}_{i}\dot{\bf R}_{j}=\frac12 ({\boldsymbol\omega}, {\bf m})
=\frac12 (RIR^T)_{ij}\omega_i\omega_j
=\frac12I_{ij}\Omega_i\Omega_j=\frac12(RI^{-1}R^T)_{ij}m_im_j=\frac12 I^{-1}_{ij}M_i M_j.
\end{eqnarray}
The mass matrix $g_{ij}$ and the  tensor of inertia $I_{ij}$, appeared in this expressions, should be calculated in the Laboratory system with the origin at the fixed point instead of the center-of-mass point. 

Further, substituting the anzatz (\ref{fp6}) into the equations of motion (\ref{fp3}) and analysing them, we arrive at the second-order  equations of motion for the rotational degrees of freedom $R_{ij}$: $\ddot R_{ik}g_{kj}=-R_{ik}\lambda_{kj}(\lambda_{A N}(t))$.
They follow from its own Lagrangian action 
\begin{eqnarray}\label{fp8}
S=\int dt ~ ~ \frac12 g_{ij}\dot R_{ki}\dot R_{kj} -\frac12 \lambda_{ij}\left[R_{ki}R_{kj}-\delta_{ij}\right]. 
\end{eqnarray}
Following the procedure of Sect. VI of the work \cite{AAD23}, we rewrite the second-order equations in the first-order form and exclude the auxiliary variables $\lambda_{ij}$. In the result,  the evolution of a rigid body with a fixed point can be described by $3+9$ equations for mutually independent variables $R_{ij}$ and $\Omega_k$
\begin{eqnarray}\label{fp9}
I\dot{\boldsymbol\Omega}=[I{\boldsymbol\Omega}, {\boldsymbol\Omega}], \label{s0} \qquad    
\dot R_{ij}=-\epsilon_{jkm}\Omega_k R_{im}.  \label{s1}
\end{eqnarray}
They should be resolved with the universal initial data $R_{ij}(0)=\delta_{ij}$ and  ${\boldsymbol\Omega}(0)={\boldsymbol\Omega}_0$, implied by Eq. (\ref{fp6}). By construction, the solutions to these equations with other initial data for $R_{ij}$ are not related to the motions of a rigid body. 

The rotation matrix $R_{ij}$ turns out to be the basic quantity of the formalism, since it contains all the information about time evolution of the body in Laboratory system, see Eq. (\ref{fp6}). 

By $I$ in Eq. (\ref{s0}) was denoted the inertia tensor. This is a numeric $3\times 3$\,-matrix defined as follows:
\begin{eqnarray}\label{s2}
I_{ij}\equiv\sum_{N=1}^{n}m_N\left[{\bf x}_N^2(0)\delta^{ij}-x_N^i(0)x_N^j(0)\right]. 
\end{eqnarray}
Generally, $I_{ij}$ is a symmetric matrix 
\begin{eqnarray}\label{s0.02}
I=\left(
\begin{array}{ccc}
I_{11} & I_{12} & I_{13} \\
I_{12} & I_{22} & I_{23} \\
I_{13} & I_{23} & I_{33} 
\end{array}\right),  
\end{eqnarray}
transforming as the second-rank tensor under rotations of the Laboratory system.  So the explicit form of the numeric matrix, that appears in equations (\ref{s0}), depends on the initial position of the body. Equivalently, it can be said that it change when we pass from one Laboratory basis to another one, related by some rotation. Let us consider two orthonormal bases related by rotation with help of numeric orthogonal matrix $U^TU=1$: ${\bf e}'_i={\bf e}_kU^T_{ki}$. Coordinates of the body's particles in these bases are related as follows: 
$x'^i=U_{ij}x^j$.
Then Eq. (\ref{s2}) implies that the matrices $I'_{ij}$ and $I_{ij}$, computed in these bases, are related by 
\begin{eqnarray}\label{s0.03}
I'_{ij}=\sum_{N=1}^{n}m_N\left[{\bf x'}_N^2(0)\delta^{ij}-x'^i_N(0)x'^j_N(0)\right]=
U_{ia}\left[\sum_N {\bf x}_N^2(0)\delta^{ab}- m_Nx_N^a(0)x_N^b(0)\right]U_{jb}=U_{ia}I_{ab}U_{jb}, \quad \mbox{or} \quad I'=UIU^T. \quad 
\end{eqnarray}
Adapting the Laboratory system with the position of the body at $t=0$, we can simplify Eqs. (\ref{s0}). Indeed, assume that at the instant $t=0$ the Laboratory axes ${\bf e}_i$ have been chosen in the direction of eigenvectors of the matrix $I_{ij}$. Then the inertia tensor in Eqs. (\ref{s0}) acquires diagonal form \cite{Shi_1977}
\begin{eqnarray}\label{s0.01}
I=\left(
\begin{array}{ccc}
I_1 & 0 & 0 \\
0 & I_2 & 0 \\
0 & 0 & I_3 
\end{array}\right). 
\end{eqnarray}
As we saw above, due to initial data$R_{ij}(0)=\delta_{ij}$, the axes ${\bf R}_i(t)$ of body-fixed basis at $t=0$ coincide with the Laboratory axes ${\bf e}_i$, and therefore coincide also with the inertia axes.  Since the axes ${\bf R}_i(t)$ and the inertia axes are rigidly connected with the body, they will coincide in all future moments of time.

Let's consider an asymmetric rigid body, that is ($I_1\ne I_2\ne I_3$), and suppose that we describe it using the equations (\ref{fp9}), in which the inertia tensor is chosen to be diagonal. This implies that the position of the Laboratory system is completely fixed, as described above. If for some reasons we want to choose a different coordinate system, we will be forced to use equations (\ref{s0}) with the symmetric matrix (\ref{s0.02}) containing non zero off-diagonal elements\footnote{Failure to take this circumstance into account leads to a lot of confusion, see \cite{AAD23_3}.} instead of diagonal matrix (\ref{s0.01}).

As will be seen further, it is precisely this circumstance that is not taken into account in textbooks when formulating the equations of heavy symmetric top and solving them.

\section{Heavy body with a fixed point.}\label{HB}
\begin{figure}[t] \centering
\includegraphics[width=12cm]{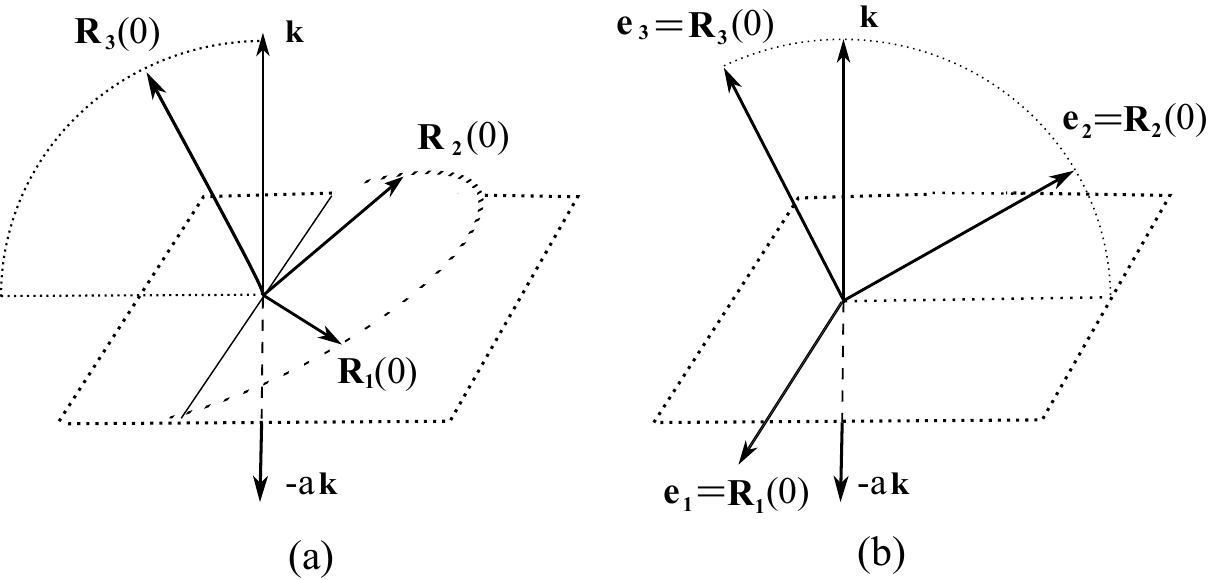}
\caption{(a) - Initial position of a heavy body with orthogonal inertia axes ${\bf R}_1(0), {\bf R}_2(0), {\bf R}_3(0)$. (b) - For the symmetric body, due to the freedom in the choice of ${\bf R}_1(0)$ and ${\bf R}_2(0)$, the vector ${\bf k}$ can be taken in the form ${\bf k}=(0, k_2, k_3).$}\label{Heavy}
\end{figure}

Consider a body with a fixed point subjected to the force of gravity, with the acceleration of gravity equal to $a>0$ and directed opposite to the constant unit vector ${\bf k}$, see Figure \ref{Heavy}(a). Then the potential energy of the body's particle ${\bf x}_N(t)$ is $am_N({\bf k}, {\bf x}_N(t))$. Summing up the potential energies of the body's points, we get the total energy 
\begin{eqnarray}\label{hb1}
U=\sum_{N=1}^N am_N({\bf k}, {\bf x}_N(t))=b({\bf k}, {\bf z}(t))=b({\bf k}, R(t){\bf z}(0)), \qquad  b\equiv aL\mu. 
\end{eqnarray}
Here $L$ is the distance from the center of mass to the fixed point, $\mu$ is the total mass of the body and ${\bf z}(0)$ is unit vector in the direction of center of mass at $t=0$. Accounting the potential energy in the action (\ref{fp2.1}), we obtain the variational problem for the heavy body. This implies the equations of motion
\begin{eqnarray}\label{hb2}
m_A\ddot{\bf x}_A=\sum_{B=1}^3 \lambda_{AB}{\bf x}_B+\sum_{\beta=4}^n \lambda_{A\beta}{\bf x}_\beta-am_A{\bf k}, \qquad 
m_\alpha\ddot{\bf x}_\alpha=\sum_{A=1}^3 \lambda_{A\alpha}{\bf x}_A-am_\alpha{\bf k}. 
\end{eqnarray}
Contrary to the free equations, we have now only two integrals  of motion. Due to torque of gravity, the components of angular momentum are not conserved 
\begin{eqnarray}\label{hb5}
\dot{\bf m}=b[{\bf k}, {\bf z}(t)]. 
\end{eqnarray}
So, the only conserved quantities are the energy and projection of angular momentum on the direction of constant vector ${\bf k}$
\begin{eqnarray}\label{hb3}
\frac{dE}{dt}=0, \qquad \mbox{where} \qquad E=\frac12 \sum_{N=1}^nm_{N}\dot{\bf x}_N^2+b({\bf k}, {\bf z}(t)),  
\end{eqnarray}
\begin{eqnarray}\label{hb4}
\frac{d}{dt}({\bf k}, {\bf m})=0, \qquad \mbox{where} \qquad {\bf m}=\sum_{N=1}^{n}m_N[{\bf x}_N, \dot{\bf x}_N]. 
\end{eqnarray}
Accounting the potential energy in the action (\ref{fp8}), we get the variational problem for the rotational degrees of freedom
\begin{eqnarray}\label{hb6}
S=\int dt ~ ~ \frac12 g_{ij}\dot R_{ki}\dot R_{kj} -\frac12 \lambda_{ij}\left[R_{ki}R_{kj}-\delta_{ij}\right]-bR_{ij}(t)k_i z_j(0). 
\end{eqnarray}
This implies second-order dynamical equations 
\begin{eqnarray}\label{hb7}
\ddot R_{ik}g_{kj}=-R_{ik}\lambda_{kj}-bk_iz_j(0).
\end{eqnarray}
Let the initial position of the inertia axes ${\bf R}_i(0)$ of the body be as shown in Figure \ref{Heavy}(a). Assuming that the Laboratory axes ${\bf e}_i$  have been chosen in the direction of the inertia axes at $t=0$, the matrices $g_{ij}$ and $I_{ij}$ in all equations acquire the diagonal form.  

Doing the calculations similar to those of Sect. V of the work \cite{AAD23}, we can exclude the auxiliary variables $\lambda_{ij}$
\begin{eqnarray}\label{hb7.1}
\lambda_{ij}=\frac{2g_i g_j}{g_i+g_j}(\dot{\bf R}_i, \dot{\bf R}_i)-\frac{g_i({\bf R}_i, {\bf F}_j)+g_j({\bf R}_j, {\bf F}_i)}{g_i+g_j}, \qquad \mbox{where} \qquad ({\bf F}_j)_i\equiv bk_i z_j(0),
\end{eqnarray}
and write closed equations of second order for $R_{ij}$. The equivalent first-order system is given then by  $3+9$  equations 
\begin{eqnarray}
I\dot{\boldsymbol\Omega}=[I{\boldsymbol\Omega}, {\boldsymbol\Omega}]+b[R^T{\bf k}, {\bf z}(0)], \label{hb8}  \qquad \qquad \\  
\dot R_{ij}=-\epsilon_{jkm}\Omega_k R_{im}, \qquad \mbox{or} \qquad \dot{\bf G}_i=-[{\boldsymbol\Omega}, {\bf G}_i],   \label{hb9}  
\end{eqnarray}
where by ${\bf G}_i$ were denoted the rows of rotation matrix $({\bf G}_i)_j=R_{ij}$. 
From the line ${\bf k}=(k_1, k_2, k_3)^T={\bf e}_ik_i={\bf e}_iR_{ij}(R^T{\bf k})_j={\bf R}_j(R^T{\bf k})_j$, we conclude that the functions $K_j(t)\equiv (R^T{\bf k})_j$ in Eq. (\ref{hb8}) are components of the vector ${\bf k}$ in the body-fixed  basis. 

For the latter use we mention the identities
\begin{eqnarray}\label{hb7.2}
R^T{\bf k}=k_i{\bf G}_i, \qquad [R^T{\bf k}, {\bf z}(0)]_i=({\bf R}_i[{\bf k}, {\bf z}(t)]).
\end{eqnarray}
They can be used to represent the torque of gravity in various equivalent forms. 

Similarly to a free body,  Euler equations (\ref{hb8}) are equivalent to the equations (\ref{hb5}). Indeed, for the angular momentum in the 
body ${\bf M}=R^T{\bf m}$ the Eq. (\ref{hb5}) implies
\begin{eqnarray}\label{hb7.3}
\dot{\bf M}=-[I^{-1}{\bf M}, {\bf M}]+b[R^T{\bf k}, {\bf z}(0)]. 
\end{eqnarray}
Since ${\bf M}=I{\boldsymbol\Omega}$, these are just the Euler equations.

{\bf Hamiltonian character of Euler-Poisson equations.} The equations (\ref{hb8}) and (\ref{hb9}) represent a Hamiltonian system \cite{Dir_1950,GT,deriglazov2010classical}. This can be confirmed by constructing the Hamiltonian formulation of the Lagrangian theory (\ref{hb6}) with help of intermediate formalism developed in the work \cite{AAD23_2}. This gives the Hamiltonian 
\begin{eqnarray}\label{hb7.3.1}
H=\frac12 I_{ij}\Omega_i\Omega_j+bR_{ij}k_iz_j(0),
\end{eqnarray}
where $\Omega_j$ are Hamiltonian counterparts of angular velocity in the body. The corresponding symplectic structure on phase 
space with the coordinates $R_{ij}, \Omega_k$ reads as follows:
\begin{eqnarray}\label{hb7.3.2}
\{R_{ij}, R_{kn}\}=0, \qquad \{\Omega_{i}, \Omega_{j}\}=-\frac{I_k}{I_i I_j}\epsilon_{ijk}\Omega_k, \qquad 
\{\Omega_{i}, R_{jk}\}=-\frac{1}{I_i}\epsilon_{ikm}R_{jm}.
\end{eqnarray}
They coincide with the brackets suggested by Chetaev \cite{Chet_1941, Chet_1989} as a symplectic structure of the theory (\ref{hb8}) and (\ref{hb9}), see \cite{AAD23} for the details.  By construction, the brackets are degenerate, and $(RR^T)_{ij}$ are their Casimir functions.
In the Hamiltonian formalism, the Euler-Poisson equations acquire the form $\dot q^A=\{ q^A, H\}$, where $q^A$ is the set of phase-space variables $R_{ij}, \Omega_k$.

{\bf Partial separation of variables in basic equations and the Euler-Poisson equations.}  Assuming $k_3\ne 0$, consider the change of variables 
\begin{eqnarray}\label{hb10.1}
({\boldsymbol\Omega}, {\bf G}_1, {\bf G}_2, {\bf G}_3)\rightarrow({\boldsymbol\Omega}, {\bf G}_1, {\bf G}_2, {\bf K}), 
\qquad \mbox{where} \qquad {\bf K}=k_i{\bf G}_i=R^T{\bf k}. 
\end{eqnarray}
Using the identities (\ref{hb7.2}) we can separate $3+3$ Euler-Poisson equations of the system (\ref{hb8}) and (\ref{hb9}) 
\begin{eqnarray}
I\dot{\boldsymbol\Omega}=[I{\boldsymbol\Omega}, {\boldsymbol\Omega}]+b[{\bf K}, {\bf z}(0)], \label{hb10}  \\   
\dot{\bf K}=-[{\boldsymbol\Omega}, {\bf K}], \qquad \qquad\label{hb11}  
\end{eqnarray}
from the remaining $3+3$  equations
\begin{eqnarray}\label{hb12}
\dot{\bf G}_1=-[{\boldsymbol\Omega}, {\bf G}_1], \qquad \dot{\bf G}_2=-[{\boldsymbol\Omega}, {\bf G}_2]. 
\end{eqnarray}
Using (\ref{fp7}), the integrals of motion (\ref{hb3}), (\ref{hb4}) can be rewritten as the integrals of motion of the system (\ref{hb10}) and (\ref{hb11}) as follows: 
\begin{eqnarray}\label{hb13}
E=\frac12 I_i\Omega_i^2+b({\bf K}, {\bf z}(0)), \qquad ({\bf k}, {\bf m})=({\bf K}, I{\boldsymbol\Omega}), \qquad {\bf K}^2=1.  
\end{eqnarray}
The equations (\ref{hb10}) and (\ref{hb11}) also form a Hamiltonian system with the Hamiltonian and brackets defined as follows:
\begin{eqnarray}\label{hb7.3.3}
H=\frac12 I_{ij}\Omega_i\Omega_j+b({\bf K}, {\bf z}(0)),
\end{eqnarray}
\begin{eqnarray}\label{hb7.3.4}
\{K_{i}, K_{j}\}=0, \qquad \{\Omega_{i}, \Omega_{j}\}=-\frac{I_k}{I_i I_j}\epsilon_{ijk}\Omega_k, \qquad 
\{\Omega_{i}, K_{k}\}=-\frac{1}{I_i}\epsilon_{ikm}K_{m}.
\end{eqnarray}
Here ${\bf K}^2$ turns out to be the Casimir function of the brackets. 

We emphasize once again that for asymmetric body $I_1\ne I_2\ne I_3$ there is no more a freedom to simplify the equations using a rotation of the  Laboratory frame. In particular, the torque in Eq. (\ref{hb10}) generally contains all three components of the vector ${\bf K}$. In component form, the equations  (\ref{hb10}) and (\ref{hb11}) read as follows
\begin{eqnarray}
\dot\Omega_1=\frac{I_2-I_3}{I_1}\Omega_2\Omega_3+\frac{b}{I_1}(K_2z_3-K_3z_2), \qquad \qquad \cr \dot\Omega_2=\frac{I_3-I_1}{I_2}\Omega_1\Omega_3+\frac{b}{I_2}(K_3z_1-K_1z_3),  \qquad \qquad \cr \dot\Omega_3=\frac{I_1-I_2}{I_3}\Omega_1\Omega_2+\frac{b}{I_3}(K_1z_2-K_2z_1),  \qquad \qquad  \label{hb14} \\  
\dot K_1=\Omega_3K_2-\Omega_2K_3, \quad \dot K_2=-\Omega_3K_1+\Omega_1K_3, \quad  \dot K_3=\Omega_2K_1-\Omega_1K_2.  \label{hb15}  
\end{eqnarray}
Their formal solution can be written in terms of exponential of the Hamiltonian vector field, see \cite{AAD23_2}.
For the center-of-mass vector ${\bf z}(0)$ in a general position, solution to these equations in quadratures is not known. There are two special cases, when the solution can be found in quadrarures: Lagrange and Kovalevskaya tops. The Lagrange top will be revised in the next section. The discussion of Kovalevskaya top \cite{Kow_1899} in modern form can be found in \cite{Per_2001}.

\section{Lagrange top.}\label{LKT} 

Let us take a symmetric body, that is $I_1=I_2\ne I_3$. Then the Euler equations (\ref{hb8}) can be simplified: without loss of generality, we can assume  that the vector ${\bf k}$ has the following form: ${\bf k}=(0, k_2, k_3)$. 
Indeed, the eigenvectors and eigenvalues of the inertia tensor $I$ obey the relations $I{\bf R}_i(0)=I_i{\bf R}_i(0)$.  With $I_1=I_2$ we have $I{\bf R}_1(0)=I_2{\bf R}_1(0)$ and $I{\bf R}_2(0)=I_2{\bf R}_2(0)$, then any linear combination $\alpha{\bf R}_1(0)+\beta{\bf R}_2(0)$ also represents an eigenvector with eigenvalue $I_2$. This means that we are free to choose any two orthogonal axes on the plane ${\bf R}_1(0), {\bf R}_2(0)$ as the inertia axes. Hence, in the case $I_1=I_2$ we can rotate the Laboratory axes in the plane $(x^1, x^2)$  without breaking the diagonal form of the inertia tensor. Using this freedom, we can assume that $k_1=0$ for our problem, see Figure \ref{Heavy}(b).  

Passing to the  case of Lagrange top, we further assume that the fixed point of the symmetric body was chosen such that center of mass lies on the third axis of inertia. Then at the initial instant of time we have ${\bf z}(0)=(0, 0, 1)^T$.  Substituting ${\bf k}=(0, k_2, k_3)^T$ and ${\bf z}(0)=(0, 0, 1)^T$ into  the Euler equations (\ref{hb8}) we get 
\begin{eqnarray}\label{hb16} 
\dot\Omega_1=\phi\Omega_2+\frac{b}{I_2}(k_2R_{22}+k_3R_{32}), \qquad \dot\Omega_2=-\phi\Omega_1-\frac{b}{I_2}(k_2R_{21}+k_3R_{31}), \qquad \dot\Omega_3=0, 
\end{eqnarray}
where $\phi=\frac{I_2-I_3}{I_2}\Omega_3=\mbox{const}$. 
Together with (\ref{hb9}), they represent equations of motion of the Lagrange top. The last equation from (\ref{hb16}) implies that besides the integrals of motion (\ref{hb3}) and (\ref{hb4}) there is one more: $\Omega_3=\mbox{const}$. This could be seen also from Eq. (\ref{hb7.3}). Indeed, the third component of this equation reads as follows: 
\begin{eqnarray}\label{hb17}
\dot M_3=\left(\frac{1}{I_2}-\frac{1}{I_1}\right)M_1 M_2-b({\bf k},[{\bf R}_3(t), {\bf z}(t)]). 
\end{eqnarray}
For the Lagrange top we have $I_1=I_2$ and ${\bf z}(t)={\bf R}_3(t)$, so $\dot M_3=I_3\dot\Omega_3=0$. 

{\bf Variational problem for the Lagrange top in terms of Euler angles.}  Let us write the Lagrangian (\ref{hb6}) as follows:
\begin{eqnarray}\label{hb24}
L= \frac12 I_i(\Omega_i)^2-\frac12 \lambda_{ij}\left[R_{ki}R_{kj}-\delta_{ij}\right]-bR_{ij}(t)k_i z_j(0), 
\end{eqnarray}
where $\Omega_i\equiv-\frac12 \epsilon_{ijk}(R^T\dot R)_{jk}$. Let us substitute the expression for $R_{ij}$ in terms of Euler angles 
\begin{eqnarray}\label{6.3}
R=\left(
\begin{array}{ccc}
\cos\psi\cos\varphi-\sin\psi\cos\theta\sin\varphi  &  -\sin\psi\cos\varphi-\cos\psi\cos\theta\sin\varphi &\sin\theta\sin\varphi \\
\cos\psi\sin\varphi+\sin\psi\cos\theta\cos\varphi & -\sin\psi\sin\varphi+\cos\psi\cos\theta\cos\varphi & -\sin\theta\cos\varphi \\
\sin\psi\sin\theta & \cos\psi\sin\theta & \cos\theta
\end{array}\right)
\end{eqnarray}
into Eq. (\ref{hb24}).  According to classical mechanics\footnote{See Sect. 17 in \cite{Arn_1} or Sect. 1.6 in \cite{deriglazov2010classical}.}, this gives an equivalent variational problem. Since the rotation matrix in terms of Euler angles authomatically obeys the  constraint $R^TR={\bf 1}$, the second term of the action (\ref{hb24}) vanishes, and we get
\begin{eqnarray}\label{hb24.1}
L=\frac12 I_2[\dot\theta^2+\dot\varphi^2\sin^2\theta]+ \frac12 I_3[\dot\psi+\dot\varphi\cos\theta]^2-bR_{ij}(\theta, \varphi, \psi)k_i z_j(0).
\end{eqnarray}
As we saw above, the Lagrange top corresponds to the choice ${\bf z}(0)=(0, 0, 1)$ and $k=(0, k_2, k_3)$. With these ${\bf z}(0)$ and ${\bf k}$ the potential energy acquires the form $b(k_2R_{23}+k_3R_{33})$, and we get the variational problem for the Lagrange top in terms of Euler angles 
\begin{eqnarray}\label{ss7}
L=\frac12 I_2[\dot\theta^2+\dot\varphi^2\sin^2\theta]+ \frac12 I_3[\dot\psi+\dot\varphi\cos\theta]^2-b[k_3\cos\theta-k_2\sin\theta\cos\varphi].
\end{eqnarray}
This gives the following equations of motion 
\begin{eqnarray}
I_3[\dot\psi+\dot\varphi\cos\theta]=m_\psi=\mbox{const},  \label{ss6.1} \\ 
\frac{d}{dt}[I_2\dot\varphi\sin^2\theta+m_\psi\cos\theta]+bk_2\sin\theta\sin\varphi=0,   \label{ss6.2} \\ 
-I_2\ddot\theta+I_2\dot\varphi^2\sin\theta\cos\theta-m_\psi\sin\theta\dot\varphi+b[k_3\sin\theta+k_2\cos\theta\cos\varphi]=0. \label{ss6.3}
\end{eqnarray}
{\bf Comment. } In many textbooks \cite{Whit_1917, Mac_1936, Lei_1965, Gol_2000, Grei_2003, Ham_22}, authors considered another equations following from a dfferent Lagrangian \cite{Arn_1,Landau_8}, the latter does not contain the term  proportional to $k_2$ 
\begin{eqnarray}\label{ss7.1}
L_A=\frac12 I_2[\dot\theta^2+\dot\varphi^2\sin^2\theta]+ \frac12 I_3[\dot\psi+\dot\varphi\cos\theta]^2-b\cos\theta. 
\end{eqnarray}
This term is discarded on the base of the following reasoning: to simplify the analysis, choose the Laboratory axis ${\bf e}_3$ in the direction of gravity  vector ${\bf k}$. However, this reasoning does not take into account the presence in the equations of moments of inertia, which have the tensor law of transformation under rotations. Indeed, going back to Eqs. (\ref{hb8}) and (\ref{hb9}), select ${\bf e}_3$ in Figure \ref{Heavy}(b) in the direction of ${\bf k}$, and calculate the components of the inertia tensor. Since the axis of inertia ${\bf R}_3(0)$ does not coincide with ${\bf e}_3$, we obtain a symmetric matrix with non-zero off-diagonal elements (\ref{s0.02}) instead of (\ref{s0.01}). This symmetric matrix should now be used to construct the kinetic part of Lagrangian and hence it appears in the equations of motion. That is, the attempt to simplify the potential energy will lead, instead of (\ref{ss7.1}), to a Lagrangian with a complicated expression for the kinetic energy. 

Does a rotating body have motions that could be described using the equations following from incomplete Lagrangian (\ref{ss7.1})? The answer is yes: these are solutions with special initial data, for which the ${\bf R}_3(t)$\,-axis coincides with ${\bf k}$ at some (finite) instant of time. These are the solutions of an awakened top and its limiting case of a sleeping top, see below. In the general case, to look for the solutions with ${\bf R}_3(t)$ that do not pass through ${\bf k}$, one should use the equations following from (\ref{ss7}).

Probably for the first time in the monographic literature the equations following from incomplete Lagrangian (\ref{ss7.1})  were discussed in details by MacMillan in \cite{Mac_1936}. In the absence of analytical solution in elementary functions, MacMillan performed analysis of integrals of motion and effective potential, reducing the problem to the study of a polynomial of degree 3. The results of this qualitative analysis are summarized in Figs. 60-62 of his book, and then reproduced in many other textbooks. In this respect we point out that a similar analysis of the improved Lagrangian (\ref{ss7}) leads to the study of a polynomial of degree 6, see \cite{AAD23_5}.

\section{\bf Sleeping Lagrange top.} 
Consider the Lagrange top that at $t=0$ has its center-of-mass vector in the direction of gravity vector ${\bf k}$, and was launched with initial angular velocity $\Omega_0$=const around the axis ${\bf k}$. In accordance with this, we use ${\bf z}(0)={\bf k}=(0, 0, 1)^T$ and ${\boldsymbol\Omega}(0)=(0, 0, \Omega_0)$ in Eqs. (\ref{hb8}), (\ref{hb9}). The equations for $\Omega(t)$ and ${\bf R}_3(t)$ read as follows
\begin{eqnarray}\label{an1}
\dot\Omega_1=\phi\Omega_2+\frac{b}{I_2}R_{32}, \quad \dot\Omega_2=-\phi\Omega_1-\frac{b}{I_2}R_{31}, \quad \dot\Omega_3=0;   \qquad \qquad \cr 
\dot R_{31}=R_{32}\Omega_3-R_{33}\Omega_2, \quad \dot R_{32}=R_{33}\Omega_1-R_{31}\Omega_3, \quad 
\dot R_{33}=R_{31}\Omega_2-R_{32}\Omega_1,   
\end{eqnarray}
where $\phi=\frac{I_2-I_3}{I_2}\Omega_3$. 
They are satisfied by the functions ${\boldsymbol\Omega}(t)=(0, 0, \Omega_0)$,  $R_{31}(t)=R_{32}(t)=0$ and $R_{33}(t)=1$. Then the remaining equations from 
(\ref{hb8}), (\ref{hb9}) are 
\begin{eqnarray}\label{an2}
\dot R_{11}=R_{12}\Omega_0, \qquad \dot R_{12}=-R_{11}\Omega_0,  \qquad \dot R_{13}=0;  \cr 
\dot R_{21}=R_{22}\Omega_0, \qquad \dot R_{22}=-R_{21}\Omega_0,  \qquad \dot R_{23}=0. 
\end{eqnarray}
Their general solution is
\begin{eqnarray}\label{an3}
R_{11}=A\cos\Omega_0 t+B\sin\Omega_0 t, \quad R_{12}=-A\sin\Omega_0 t+B\cos\Omega_0 t, \quad R_{13}=F; \cr
R_{21}=C\cos\Omega_0 t+D\sin\Omega_0 t, \quad R_{22}=-C\sin\Omega_0 t+D\cos\Omega_0 t, \quad R_{23}=G. 
\end{eqnarray}
Taking into account the initial data $R_{ij}(0)=\delta_{ij}$, we get $A=D=1$, $B=C=F=G=0$, and the solution is the stationary rotation around the vector of gravity ${\bf k}$
\begin{eqnarray}\label{an4}
R_{ij}(t)=\left(
\begin{array}{ccc}
\cos\Omega_0 t & -\sin\Omega_0 t & 0 \\
\sin\Omega_0 t & \cos\Omega_0 t & 0 \\
0 & 0 & 1
\end{array}
\right). 
\end{eqnarray}

\section{Awakened Lagrange top.} 
Consider the Lagrange top that at $t=0$ has its center-of-mass vector in the direction of gravity vector ${\bf k}$, so ${\bf z}(0)={\bf k}=(0, 0, 1)^T$. Substituting these values into Eq. (\ref{hb24}) we get the following variational problem in terms of Euler angles
\begin{eqnarray}\label{an5}
L=\frac12 I_2[\dot\theta^2+\dot\varphi^2\sin^2\theta]+ \frac12 I_3[\dot\psi+\dot\varphi\cos\theta]^2-b\cos\theta.
\end{eqnarray}
The initial data should be formulated now for the Euler angles. Let the top was lounched from vertical position with some linear velocity $\dot {\bf R}_3(0)$.  This vector is parallel to the plane of the Laboratory vectors ${\bf e}_1$ and ${\bf e}_2$.  Using the freedom to rotate the Laboratory system in the plane of these vectors without spoiling the diagonal form of the inertia tensor, we can assume that $\dot {\bf R}_3(0)$ is antiparallel to ${\bf e}_2$. With these agreements, consider the top with the initial position $\theta(0)=\varphi(0)=\psi(0)=0$, and with the initial nutation $\dot\theta(0)=\dot\theta_0$ and rotation $\dot\psi(0)=\dot\psi_0$, where $\dot\theta_0<<\dot\psi_0$. We do not fix the speed of precession $\dot\varphi(0)$ of azimyth plane  because it is determined by $\dot\psi_0$, see Eq. (\ref{an10}) below. The Lagrangian (\ref{an5}) does not depend on $\psi$ and $\varphi$, so the equations of motion $\delta S/\delta\psi=0$ and $\delta S/\delta\varphi=0$ give the integrals of motion $m_\psi$ and $m_\varphi$
\begin{eqnarray}
I_3[\dot\psi+\dot\varphi\cos\theta]=m_\psi=\mbox{const}, \qquad \mbox{then} \quad m_\psi=I_3[\dot\psi_0+\dot\varphi_0], \label{an6} \\
I_2\dot\varphi\sin^2\theta+m_\psi\cos\theta=m_\varphi=\mbox{const}.  \qquad \qquad \label{an7} 
\end{eqnarray}
Substituting the initial data $\theta(0)=0$ into the last equation, we conclude that the integrals of motion are not independent
\begin{eqnarray}\label{an8}
m_\varphi=m_\psi.
\end{eqnarray}
Taking this into account, Eq. (\ref{an7} ) reads
\begin{eqnarray}\label{an9}
\dot\varphi=\frac{m_\psi}{I_2(1+\cos\theta)}. 
\end{eqnarray}
This implies that $\dot\varphi$ is a non negative function, so the azimuth plane can not change its direction of rotation during the motion of the top. Besides, from the above expressions we get the initial speed $\dot\varphi_0$ and then the integral of motion  $m_\psi$ through the initial rotation $\dot\psi_0$
\begin{eqnarray}\label{an10}
\dot\varphi_0=\frac{I_3}{2I_2-I_3}\dot\psi_0,  \qquad m_\psi=\frac{2I_2 I_3}{2I_2-I_3}\dot\psi_0. 
\end{eqnarray}
Variation of the Lagrangian (\ref{an5}) with respect to $\theta$ gives the second-order equation
\begin{eqnarray}\label{an11}
-I_2\ddot\theta+I_2\dot\varphi^2\sin\theta\cos\theta-I_3[\dot\psi+\dot\varphi\cos\theta]\dot\varphi\sin\theta+b\sin\theta=0.
\end{eqnarray}
Using Eqs. (\ref{an6}) and (\ref{an10}), we exclude the variables $\varphi$ and $\psi$, obtaining closed equation for $\theta$
\begin{eqnarray}\label{an12}
-I_2\ddot\theta-\frac{m_\psi^2}{I_2}\frac{\sin\theta}{(1+\cos\theta)^2}+b\sin\theta=0. 
\end{eqnarray}
This equation follows from the effective Lagrangian 
\begin{eqnarray}\label{an13}
L_{eff}=\frac{I_2}{2}\dot\theta^2-\frac{m_\psi^2}{I_2(1+\cos\theta)}-b\cos\theta.
\end{eqnarray}
The energy of this effective one-dimensional problem is an integral of motion 
\begin{eqnarray}\label{an14}
\frac{I_2}{2}\dot\theta^2+\frac{m_\psi^2}{I_2(1+\cos\theta)}+b\cos\theta=E=\mbox{const}, \qquad \mbox{then} \quad E=\frac12[I_2\dot\theta_0^2+\frac{m_\psi^2}{I_2}]+b.
\end{eqnarray}
This allows us to write the first-order equation for $\theta$
\begin{eqnarray}\label{an15}
\dot\theta=\sqrt{\frac{2}{I_2}}\sqrt{\frac{(E-b\cos\theta)(1+\cos\theta)-\gamma}{1+\cos\theta}}, \qquad \mbox{where} \quad 
\gamma\equiv\frac{m_\psi^2}{I_2}, 
\end{eqnarray}
that can be immediately integrated 
\begin{eqnarray}\label{an16}
\int\frac{dx}{\sqrt{x[(E-b+bx)(2-x)-\gamma]}}=\sqrt{\frac{2}{I_2}}\int dt, \qquad \mbox{where} \quad x\equiv 1-\cos\theta. 
\end{eqnarray}
So the problem was reduced to the elliptic integral appeared in the last expression. 

\begin{figure}[t] \centering
\includegraphics[width=08cm]{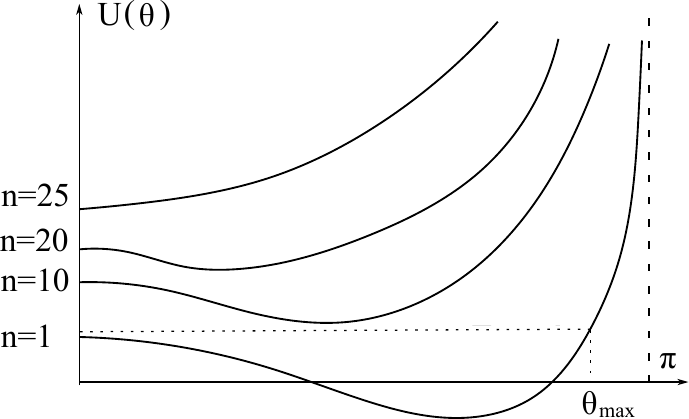}
\caption{Effective potential energy of awakened Lagrange top with initial nutation rate $\dot\theta_0=11{}^\circ$/sec.}\label{PotE}
\end{figure}

In the absence of an analytic solution in elementary functions, we can use the effective one-dimensional problem (\ref{an13}) for qualitative analysis of the motion. Consider a top of mass $\mu=0,1$ kg, in the form of a cone of height $h=0,12$ m and radius $r=0,06$ m. As the fixed point we take the vertex of the cone. Then the distance to the center of mass and the inertia moments are \cite{Landau_8}
\begin{eqnarray}\label{an17}
L=\frac{3h}{4}=0,09 m, \qquad I_1=I_2=\frac{3}{5}\mu(\frac{r^2}{4}+h^2)=918\times 10^{-6},  \qquad 
I_3=\frac{3}{10}\mu r^2=108\times 10^{-6}. 
\end{eqnarray}
As the initial velocities of rotation  and nutation  we take $\dot\psi_0=2\pi n ~ rad/sec$ and $\dot\theta_0=0,2 ~ rad/sec\sim 11^{\circ}/sec$,
where $n$ is the number of revolutions per second. The potential energy of the effective problem (\ref{an13}) reads
\begin{eqnarray}\label{an18}
U_{eff}(\theta)=\frac{\gamma}{1+\cos\theta}+b\cos\theta, \quad \mbox{where}  \quad \gamma\equiv\frac{m_\psi^2}{I_2}=0,565694455\times 10^{-3}, \quad b=a\mu L=0,0882. 
\end{eqnarray}
Typical graphs of the function $U_{eff}(\theta)$ are drawn in the figure \ref{PotE}. 
The potential energy of a slow top has a minimum at the point $\cos\theta=-1+\sqrt{\gamma/b}$, which shifts to the left with increasing the rotational speed $\dot\psi_0$, so that for velocities greater than $24$ revolutions per second the point of minimum becomes $\theta=0$. The graphs show that the awakened top first deviates from the vertical position, and then returns back.

\begin{figure}[t] \centering
\includegraphics[width=04cm]{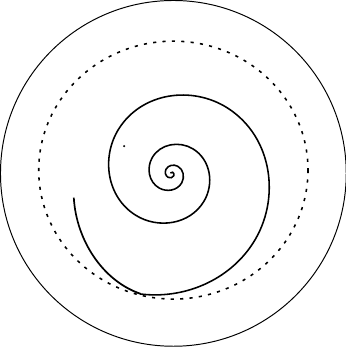}
\caption{Trajectory of the axis ${\bf R}_3(t)$ (from vertical to the maximum deviation position) of awakened Lagrange top with initial nutation rate $\dot\theta_0=11{}^\circ$/sec.}\label{Awak_1}
\end{figure}

The maximum deviation $\theta_{max}$ of the  axis ${\bf R}_3(t)$ from the vertical can be found in analytical form from the 
equation (\ref{an14}), in which we should  put $\dot\theta(t)=0$. Then
\begin{eqnarray}\label{an19}
\cos\theta_{max}=\frac{\gamma+\sigma}{4b}-\sqrt{(1+\frac{\gamma+\sigma}{4b})^2-\frac{\gamma}{b}}, \quad \mbox{where} \quad \delta=I_2\dot\theta_0^2=0,3672\times 10^{-4}.
\end{eqnarray}
Besides, with use  of Eqs. (\ref{an9}) and (\ref{an10}) we can estimate the character of rotation of the azimyth plane by calculating the precession 
speed $\dot\varphi_0$ and $\dot\varphi(\theta_{max})$. The results of these calculations are presented in the table \ref{Table_1}. The initial precession speed $\dot\varphi_0$ of the azimuth plane grows with $\dot\psi_0$, while the precession speed $\dot\varphi(\theta_{max})$ at the point of maximum deviation firstly decreases but then begins to increase, starting from 25 revolutions per second. The precession rate of a fast top changes slowly with time. Typical trajectory of the third axis ${\bf R}_3(t)$ is drawn in the Figure \ref{Awak_1}. This should be compared with Figures 60-62 of 
MacMillan's book \cite{Mac_1936}. 
\begin{center}
\begin{table}
\caption{Maximum deviation and precession rates of awakened Lagrange top with initial nutation rate $\dot\theta_0=11{}^\circ$/sec.}
\label{Table_1}
\begin{center}
\begin{tabular}{c|c|c|c}
$\dot\psi_0=n$ rev./sec ~ & ~ $\theta_{max}{}^{\circ}$ ~  &  ~ $\dot\varphi_0$ rev./sec ~ & ~ $\dot\varphi(\theta_{max})$ rev./sec \\  
\hline \hline
1   & 175 & 0,0625 & 39 \\
10 & 131 & 0,625  & 3,9 \\
13 & 117 & 0,8 & 3,0 \\
16 & 103 & 1   & 2,44 \\
20 &  74  & 1,25 & 1, 95 \\
23 & 46 & 1, 43 & 1,7 \\
25 & 11 & 1,56 & 1,58 \\
26 & 4  &  1,62 & 1,62 \\
\end{tabular}
\end{center}
\end{table}
\end{center}

\section{Horizontal precession around of gravity vector.} 
\begin{figure}[t] \centering
\includegraphics[width=10cm]{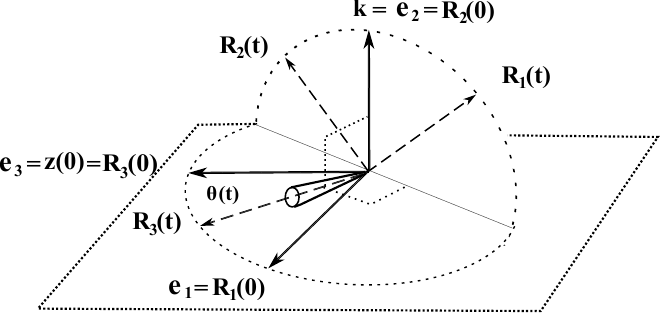}
\caption{Horizontal precession of the Lagrange top around of gravity vector.}\label{Horiz}
\end{figure}

Consider the Lagrange top that at $t=0$ has its third axis ${\bf R}_3(0)$ orthogonal to the gravity vector ${\bf k}$, see Figure \ref{Horiz}. 
Using the freedom of rotation of the Laboratory axes in the plane ${\bf e}_1, {\bf e}_2$, we can assume 
that ${\bf k}={\bf e}_2=(0, 1, 0)^T$. Substituting these values together with ${\bf z}(0)=(0, 0, 1)^T$ into the Lagrangian (\ref{hb24.1}), we get
\begin{eqnarray}\label{an20}
L=\frac12 I_2[\dot\theta^2+\dot\varphi^2\sin^2\theta]+ \frac12 I_3[\dot\psi+\dot\varphi\cos\theta]^2+b\sin\theta\cos\varphi.
\end{eqnarray}
This implies the equations
\begin{eqnarray}\label{an21}
I_3[\dot\psi+\dot\varphi\cos\theta]=m_\psi=\mbox{const}, \cr 
\frac{d}{dt}[I_2\dot\varphi\sin^2\theta+m_\psi\cos\theta]+b\sin\theta\sin\varphi=0,   \cr 
-I_2\ddot\theta+I_2\dot\varphi^2\sin\theta\cos\theta-m_\psi\sin\theta\dot\varphi+b\cos\theta\cos\varphi=0. 
\end{eqnarray}
We assume that our top has some initial rotation and precession rates $\dot\psi(0)=\dot\psi_0$ and $\dot\theta(0)=\dot\theta_0$. Then the initial position of the top is $\theta(0)=0, \varphi(0)=\frac{\pi}{2}, \psi(0)=\frac{3\pi}{2}$. Let us look for a solution of the form $\theta(t), \varphi(t)=\frac{\pi}{2}, \psi(t)$, then the equations of motion read
\begin{eqnarray}\label{an22}
\dot\psi=\dot\psi_0, \qquad \dot\theta=\frac{b}{I_3\dot\psi_0}, \qquad \ddot\theta=0, 
\end{eqnarray}
and we get the solution $\theta(t)=\frac{b}{I_3\dot\psi_0} t, \varphi(t)=\frac{\pi}{2}, \psi(t)=\dot\psi_0 t+\frac{3\pi}{2}$. Substituting these functions into Eq. (\ref{6.3}) we obtain the rotation matrix which turns out to be the composition of two rotations: counterclockwise around ${\bf e}_2$ axis  and clockwise around ${\bf e}_3$ axes
\begin{eqnarray}\label{an23}
R(t)=\left(
\begin{array}{ccc}
\cos\dot\psi_0 t \cos\dot\theta_0 t & -\sin\dot\psi_0 t\cos\dot\theta_0 t & \sin\dot\theta_0 t \\
\sin\dot\psi_0 t & \cos\dot\psi_0 t &  0  \\
-\cos\dot\psi_0 t \sin\dot\theta_0 t & \sin\dot\psi_0 t\sin\dot\theta_0 t & \cos\alpha t 
\end{array}\right)= 
\left(
\begin{array}{ccc}
\cos\dot\theta_0  t  & 0 & \sin\dot\theta_0 t  \\ 
0 & 1 & 0 \\
-\sin\dot\theta_0 t &  0 &\cos\dot\theta_0 t 
\end{array}
\right)\times \left(
\begin{array}{ccc}
\cos\dot\psi_0 t & -\sin\dot\psi_0 t & 0 \\
\sin\dot\psi_0 t & \cos\dot\psi_0 t &  0  \\
0 & 0 & 1
\end{array}
\right)
\end{eqnarray}
where 
\begin{eqnarray}\label{an24}
\dot\theta_0=\frac{b}{I_3\dot\psi_0}. 
\end{eqnarray}
Thus the Lagrange top lounched with the rotation rate $\dot\psi_0$ and precession rate $\dot\theta_0=\frac{b}{I_3\dot\psi_0}$ will precess around ${\bf k}$ in the horizontal plane. A slow spinning top must precess at high speed to stay on this plane. A fast spinning top precesses slowly. The relationship between frequencies of rotation and precession does not depend on the geometry of the top.

\section{Inclined Lagrange top: precession around of gravity vector without nutation.} 
\begin{figure}[t] \centering
\includegraphics[width=07cm]{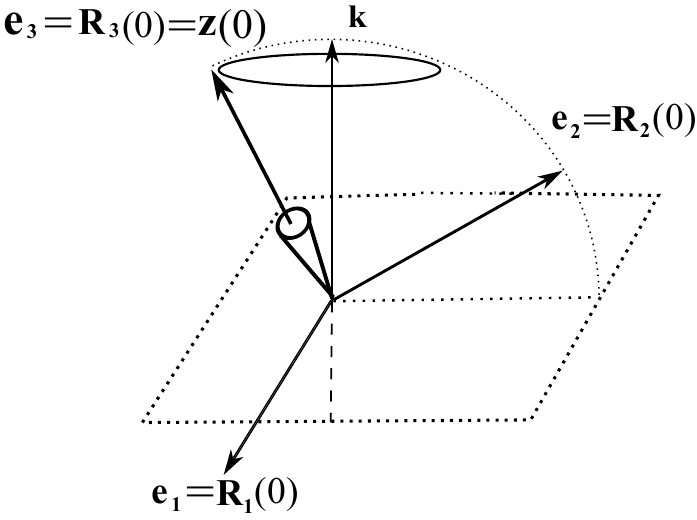}
\caption{Precession without nutation of inclined Lagrange top.}\label{Inclin}
\end{figure}

Without loss of generality, we can choose the initial position of inclined Lagrange top (and hence the Laboratory system) as shown in the Figure \ref{Inclin}. 
According (\ref{hb16}) and (\ref{hb9}), the equations of motion are
\begin{eqnarray}\label{an26} 
\dot\Omega_1=\phi\Omega_2+\frac{b}{I_2}(k_2R_{22}+k_3R_{32}), \qquad \dot\Omega_2=-\phi\Omega_1-\frac{b}{I_2}(k_2R_{21}+k_3R_{31}), \qquad \Omega_3=\mbox{const}; 
\end{eqnarray}
\begin{eqnarray}\label{an27}  
\dot R_{ij}=-\epsilon_{jkm}\Omega_k R_{im},  
\end{eqnarray}
where $\phi=\frac{I_2-I_3}{I_2}\Omega_3=\mbox{const}$.  We look for the solution that represents precession around of ${\bf k}$ without nutation, that is  $({\bf R}_3(t), {\bf k})=\mbox{const}$ for any $t$.  It can be expected that this motion be described by a rotation matrix consisting of the product of rotations around the axes ${\bf k}$ and ${\bf e}_3$:  $R(t, \alpha, \gamma)=R_{{\bf k}}(t, \gamma)\times R_{{\bf e}_3}(t, \alpha)$.  Therefore, we will look for a solution in the following form (see \cite{AAD23} for the details):
\begin{eqnarray}\label{an28}
R_{ij}=\left(
\begin{array}{ccc}
\cos \gamma t\cos\alpha t-k_3\sin \gamma t\sin\alpha  t & -\cos \gamma t\sin\alpha  t-k_3\sin \gamma t\cos\alpha  t  &  k_2\sin \gamma t  \\
{} & {} & {} \\
k_3 \sin \gamma t\cos\alpha  t+(k_2^2 +k_3^2\cos \gamma t)\sin\alpha  t & 
-k_3 \sin \gamma t\sin\alpha  t+(k_2^2 +k_3^2\cos \gamma t)\cos\alpha  t & k_2k_3(1-\cos \gamma t) \\
{} & {} & {} \\
-k_2 \sin \gamma t\cos\alpha  t+k_2 k_3(1-\cos \gamma t)\sin\alpha  t &
k_2 \sin \gamma t\sin\alpha  t+k_2 k_3(1-\cos \gamma t)\cos\alpha  t & k_3^2 +k_2^2\cos \gamma t
\end{array}\right) 
\end{eqnarray}
with some frequences of rotation $\alpha$ and precession $\gamma$. For the positive values of the frequences,  $R_{{\bf e}_3}(t, \alpha)$ ia a clockwise rotation while $R_{{\bf k}}(t, \gamma)$ is a counter-clockwise.  Substituting this matrix into Eqs. (\ref{an26}) we get
\begin{eqnarray}\label{an29} 
\dot\Omega_1=\phi\Omega_2+\frac{bk_2}{I_2}\cos\alpha t, \qquad \dot\Omega_2=-\phi\Omega_1-\frac{bk_2}{I_2}\sin\alpha t. 
\end{eqnarray}
The general solution to this system with two integration constants $c$ and $\phi_0$ is  $\Omega_1=c\sin(\alpha t+\phi_0)$, $\Omega_2=c\cos(\alpha t+\phi_0)$, where 
\begin{eqnarray}\label{an30} 
\alpha=\phi+\frac{bk_2}{I_2 c}=\frac{I_2-I_3}{I_2}\Omega_3+\frac{bk_2}{I_2 c}. 
\end{eqnarray}
The initial conditions $R_{ij}(0)=\delta_{ij}$ imply $\phi_0=0$, so finally 
\begin{eqnarray}\label{an31} 
\Omega_1=c\sin\alpha t, \qquad \Omega_2=c\cos\alpha t. 
\end{eqnarray}
The functions (\ref{an31}) and (\ref{an28}) with $\alpha$ given in (\ref{an30}) satisfy the Euler equations (\ref{an26}) for arbitrary values of the constants $\Omega_3$, $c$ and $\gamma$. Let's try to choose them so that our functions also satisfy the Poisson equations (\ref{an27}). At $t=0$ the equations (\ref{an27}) turn into $\Omega_k(0)=-\frac12\epsilon_{kij}\dot R_{ij}(0)$. Substituting our functions (\ref{an31}) and (\ref{an28}) into these equalities we obtain the numbers $c$ and $\Omega_3$ through $\gamma$ as follows:
\begin{eqnarray}\label{an32}
c=k_2\gamma, \qquad \Omega_3=\frac{b}{I_3\gamma}+\frac{I_2 k_3}{I_3}\gamma. 
\end{eqnarray}
Together with  Eq. (\ref{an30}), these expressions give the following relationship between the two frequences of our problem: 
\begin{eqnarray}\label{an33} 
\alpha=\frac{A}{\gamma}+Ck_3\gamma, \qquad \mbox{where} \quad A\equiv\frac{b}{I_3}=\frac{a\mu L}{I_3}>0, \quad C\equiv\frac{I_2-I_3}{I_3}.  
\end{eqnarray}
It is verified by direct calculations that the functions (\ref{an31}) and (\ref{an28}) with these 
values of $c$, $\Omega_3$ and $\alpha$ satisfy the Poisson equations (\ref{an27}) for any value of the precession frequency $\gamma$. Hence the rotation matrix (\ref{an28}) with the rotation and precession frequences related according Eq. (\ref{an33}) describes the precession without nutation of inclined Lagrange top. Note that for the horizontal top, $k_2=1$ and $k_3=0$, the solution obtained turn into (\ref{an23}).  

The function $\alpha(\gamma)$ depends on the top's configuration $C$ and inclination $k_3$. 
To discuss this function, consider a conical top of height $h$, radius $r$ and the precession frequency $\gamma>0$.  We get (see Eq. (\ref{an17}) 
$I_2-I_3=\frac{3\mu}{20}[4h^2-r^2]$. So $C>0$ when $h>r/2$ (high top), $C=0$ when $h=r/2$ (totally symmetric top, $I_1=I_2=I_3$ ), and $C<0$ when $h<r/2$ (low top).

Then the following cases arise. 

\begin{figure}[t] \centering
\includegraphics[width=09cm]{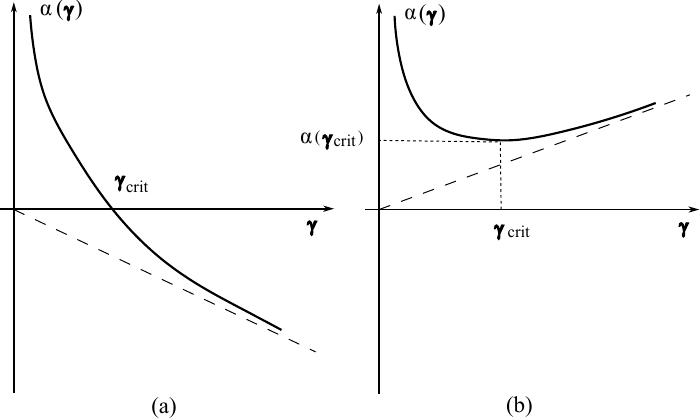}
\caption{The relationship between two frequences. (a) Low top located above the horizon, (b) High top located above the horizon.}\label{frequency}
\end{figure}

{\bf A.} The low top located above the horizon, that is 
\begin{eqnarray}\label{an34} 
2h<r, \qquad \mbox{then} \quad C<0; \qquad k_3>0.
\end{eqnarray}    
The graph of the function $\alpha(\gamma)$ is drawn in Figure \ref{frequency}(a), and implies the following behavior of the top.

\noindent {\bf A1.} Slow precession without nutation, $\gamma<\gamma_{crit}=\sqrt{-b/(I_2-I_3)k_3}$,  implies a clockwise rotation of the top around ${\bf R}_3(t)$. \par

\noindent {\bf A2.} When $\gamma=\gamma_{crit}$, the centripetal acceleration balances the force of gravity, and the top precesses without rotation around ${\bf R}_3(t)$. \par

\noindent {\bf A3.} Rapid precession, $\gamma>\gamma_{crit}$, implies counter-clockwise rotation around ${\bf R}_3(t)$. In the limit $\gamma\rightarrow\infty$ the rotation frequency $|\alpha|$  grows linearly with $\gamma$. 

{\bf B.} The high top located below the horizon 
\begin{eqnarray}\label{an35} 
2h>r, \qquad \mbox{then} \quad C>0; \qquad k_3<0,
\end{eqnarray}
has a similar bevavior. 

{\bf C.} The high top located above the horizon 
\begin{eqnarray}\label{an36} 
2h>r, \qquad \mbox{then} \quad C>0; \qquad k_3>0.
\end{eqnarray}
The graph of the function $\alpha(\gamma)$ is drawn in Figure \ref{frequency}(b), and implies the following behavior of the top.

\noindent {\bf C1.} The top rotating around ${\bf R}_3(t)$ with the frequency less than $\alpha(\gamma_{crit})=2\sqrt{(I_2-I_3)bk_3}/I_3$ cannot precess without a nutation.

\noindent {\bf C2.} There is only one rotation frequency $\alpha(\gamma_{crit})$ at which the top's precession frequency is $\gamma_{crit}=\sqrt{b/(I_2-I_3)k_3}$.

\noindent {\bf C3.} For each $\alpha>\alpha(\gamma_{crit})$, there are two possible precession frequences for the movement without nutation
\begin{eqnarray}\label{an37} 
\gamma= \frac{\alpha\pm\sqrt{\alpha^2-4ACk_3}}{2Ck_3}. 
\end{eqnarray}

{\bf D.} The low top located below the horizon 
\begin{eqnarray}\label{an38} 
2h<r, \qquad \mbox{then} \quad C<0; \qquad k_3<0,
\end{eqnarray}
has a similar bevavior. 

{\bf E.} The totally symmetrical conical top  
\begin{eqnarray}\label{an39} 
2h=r, \qquad \mbox{then} \qquad I_1=I_2=I_3, \qquad  C=0, \qquad \alpha=\frac{b}{I_3\gamma},
\end{eqnarray}
has the behavior similar to the horizontal top.  The relationship between two frequencies does not depend on the top's inclination $k_3$.

\section{Conclusion.}

When formulating and solving the equations of motion of a rigid body with the inertia tensor chosen in the diagonal form, one should keep in mind the tensor law of transformation of the moments of inertia $I_1, I_2, I_3$ under rotations. We observed that for the Lagrange top this leads to the potential energy that depends on the Euler angles $\varphi$ and $\theta$. The potential energy (see the last term in (\ref{ss7})) is different from that assumed in textbooks (see the last term in (\ref{ss7.1})). 
As far as I know, this drawback has not yet been noticed and corrected in the literature. So we revised the problem of the motion of a Lagrange  top and corrected this drawback. It should be noted that similar inaccuracies also exist when discussing a free asymmetric body, see Appendix in the recent  work \cite{AAD_2023_13}.

The problem of finding a general solution to the improved equations (\ref{ss6.1})-(\ref{ss6.3}) can be reduced to the calculation of four  elliptic integrals \cite{AAD23_5}. However, under some special initial conditions, one can either find analytical solutions in elementary functions or perform a qualitative analysis of the motion. We have found such solutions to the improved equations. The motions of an awakened and horizontally precessing Lagrange tops were analysed with use of unconstrained variables (Euler angles). The sleeping  and inclined Lagrange tops were analysed in terms of original variables (rotation matrix $R_{ij}$ and angular velocity in the body $\Omega_i$). Perhaps somewhat unexpected is the motion of a high inclined top (the case {\bf C3}): for a given rotation frequency greater than the critical one, it can precess without nutation at two different precession frequencies, see Eq. (\ref{an37}).

\begin{acknowledgments}
The work has been supported by the Brazilian foundation CNPq (Conselho Nacional de Desenvolvimento Cient\'ifico e Tecnol\'ogico - Brasil). 
\end{acknowledgments}

\end{document}